%
%
%
%
%

\newcount\chno \chno=0
\newcount\equno 
\newcount\refno \refno=0

\font\chhdsize=cmbx12 at 14.4pt

\def\startbib{\def\biblio{\bigskip\medskip
  \noindent{\chhdsize References.}\bgroup\parindent=2em}}
\def\endbib{\edef\biblio{\biblio\egroup}}
\def\reflbl#1#2{\global\advance\refno by 1
  \edef#1{\number\refno}
    \global\edef\biblio{\biblio\medskip\item{[\number\refno]}#2\par}}

\def\eqlbl#1{\global\advance\equno by 1
  \global\edef#1{{\number\chno.\number\equno}}
  (\number\chno.\number\equno)}

\def\qed{\hfil\hbox to 0pt{}\ \hbox to 2em{\hss}\ 
         \hbox to 0pt{}\hskip-2em plus 1fill
         \vrule height6pt depth1pt width7pt\par\medskip}
\def\eqed{\hfil\hbox to 0pt{}\ \hbox to 2em{\hss}\ 
          \hbox to 0pt{}\hskip-2em plus 1fill
          \vbox{\hrule height .25pt depth 0pt width 7pt
            \hbox{\vrule height 6.5pt depth 0pt width .25pt
              \hskip 6.5pt\vrule height 6.5pt depth 0pt width .25pt}
            \hrule height .25pt depth 0pt width 7pt}\par\medskip}


\def\msimp#1#2{#1%
  \hbox to 0pt{\hskip 0pt minus 3fill
    \phantom{$#1$}\hbox to 0pt{\hss$#2$\hss}\phantom{$#1$}%
    \hskip 0pt minus 1fill}}


\def\pr{{\prime}}
\def\ppr{{\prime\prime}}

\def\({\left(}
\def\){\right)}
\def\[{\left[}
\def\]{\right]}
\def\<{\left\langle}
\def\>{\right\rangle}

\def\eps{\epsilon}
\def\veps{\varepsilon}

\font\tendouble=msbm10 \font\sevendouble=msbm7  
\font\fivedouble=msbm5

\newfam\dbfam
\textfont\dbfam=\tendouble \scriptfont\dbfam=\sevendouble
\scriptscriptfont\dbfam=\fivedouble

\mathchardef\dbA="7041 
\mathchardef\dbB="7042 
\mathchardef\dbC="7043 \def\CC{{\fam=\dbfam\dbC}}
\mathchardef\dbD="7044 
\mathchardef\dbE="7045 
\mathchardef\dbF="7046 
\mathchardef\dbG="7047 
\mathchardef\dbH="7048 
\mathchardef\dbI="7049 
\mathchardef\dbJ="704A 
\mathchardef\dbK="704B 
\mathchardef\dbL="704C 
\mathchardef\dbM="704D 
\mathchardef\dbN="704E \def\NN{{\fam=\dbfam\dbN}}
\mathchardef\dbO="704F 
\mathchardef\dbP="7050 
\mathchardef\dbQ="7051 
\mathchardef\dbR="7052 \def\RR{{\fam=\dbfam\dbR}}
\mathchardef\dbS="7053 \def\SS{{\fam=\dbfam\dbS}}
\mathchardef\dbT="7054 
\mathchardef\dbU="7055 
\mathchardef\dbV="7056 
\mathchardef\dbW="7057 
\mathchardef\dbX="7058 
\mathchardef\dbY="7059 
\mathchardef\dbZ="705A

\def\bx{{\bf x}}
\def\by{{\bf y}}

\def\Im{{\cal I}m}

\def\cE{{\cal E}}
\def\cF{{\cal F}}

\def\cS{{\cal S}}


\magnification= 1200 
\baselineskip= 4pt


\startbib

\reflbl\adlerA{
	Adler, B., and von Moerbeke, P.,
		{\it Matrix integrals, Toda symmetries, Virasoro constraints,
	and orthogonal polynomials},
	Duke Math. J. {\bf 80}, pp. 863--911 [1995].}

\reflbl\adlerB{
	Adler, B., Shiota, T., and von Moerbeke, P.
		{\it Random matrices, vertex operators, and the Virasoro 
	algebra},
	Phys. Lett. {\bf A 208}, pp. 67--78 [1995].}

\reflbl\altschuleretal{
	Altschuler, E.L., Williams, T.J., Ratner, E.R., Tipton, R.,
	Stong, R., Dowla, F., and Wooten, F.,
		{\it Possible global minimum lattice configurations for
	Thomson's problem of charges on the sphere,}
	Phys. Rev. Lett. {\bf 78}, pp. 2681--2685 [1997].}

\reflbl\bleherits{
	Bleher, P., and Its, A.R., 
		{\it Asymptotics of orthogonal polynomials
	and universality in matrix models},
	Preprint (University of Texas, Austin, electronic archives,  
	Math. Phys. \# 97-85), subm. to Annals Math. }

\reflbl\boutetetal{
	Boutet de Monvel, A., Pastur, L., and Shcherbina, M.,
		{\it On the statistical mechanics approach in the random
	matrix theory: integrated density of states,}
	J. Stat. Phys. {\bf 79}, pp. 585--611 [1995].}

\reflbl\clmpCMPcan{
	Caglioti, E., Lions, P. L., Marchioro, C., and Pulvirenti, M., 
		{\it A special class of stationary flows for two-dimensional
        Euler equations: A statistical mechanics description}, 
        Commun. Math. Phys. {\bf 143}, pp. 501--525 [1992].}

\reflbl\caillol{
	Caillol, J.M.,
		J. Phys. Lett. (Paris) {\bf 46}, pp. L-245, [1981].}

\reflbl\costinjll{
	Costin, O., and Lebowitz, J.L.,
		{\it Gaussian fluctuations in random matrices},
	Phys. Rev. Lett. {\bf 75}, pp. 69--72 [1995].}

\reflbl\deiftetal{
	Deift, P., Kriecherbauer, T., McLaughlin, K. T.-R., Venakides, S., 
	and Zhou, X., 
		{\it Asymptotics for Polynomials Orthogonal with Respect 
	to Varying Exponential Weights}, 
	Int. Math. Res. Not. {\bf 16}, pp. 759--782 [1997].}

\reflbl\difrancesco{
    DiFrancesco, P., Grinsparg, P., and Zinn-Justin, J.,
	    {\it 2D Quantum Gravity and Random Matrices},
    Phys. Rep. C {\bf 254}, pp. 1--133 [1995].}

\reflbl\durrettBOOK{
	Durrett, R., 
		{\it Probability: Theory and Examples},
	$2^{nd}$ ed., Duxbury Press [1995].}

\reflbl\dynkin{
	Dynkin, E. B., 
		{\it Klassy ekvivalentnyh slu$\check{c}$a$\check{i}$nyh 
	veli$\check{c}$in}, 
        Uspeki Mat. Nauk. {\bf 6}, pp. 125--134 [1953].}

\reflbl\dysonA{	
	Dyson, F. J.,	 
		{\it Statistical theory of the energy levels 
	of a complex system,} 
	{\it Part I}, 	J. Math. Phys. {\bf 3}, pp. 140--156 [1962]; 
	{\it Part II}, 	ibid. pp. 157--165; 
	{\it Part III}, ibid. pp. 166--175.}

\reflbl\dysonB{
	Dyson, F. J.,
		{\it A Brownian motion model for the eigenvalues 
	of a random matrix,}
	J. Math. Phys. {\bf 3}, pp. 1191--1198 [1962].}

\reflbl\ellisBOOK{
	Ellis, R. S.,
		{\it Entropy, large deviations, and statistical mechanics}, 
	Springer-Verlag, New York [1985].}

\reflbl\finetti{
	de Finetti, B., 
		{\it Funzione caratteristica di un fenomeno aleatorio}, 
	Atti della R. Accad. Naz. dei Lincei, Ser. 6, Memorie, 
	Classe di Scienze Fisiche, Matematiche e Naturali 4,
        pp. 251--300 [1931].}

\reflbl\foglershklovskii{
	Fogler, M. M., and Shklovskii, B. I.,
		{\it Probability of an eigenvalue fluctuation in an interval
	of a random matrix spectrum}, 
	Phys. Rev. Lett. {\bf 74}, pp. 3312--3315 [1995].}

\reflbl\forrester{
	Forrester, P.J., 
		{\it Fluctuation formula for complex random matrices},
	Preprint, Univ. Melbourne [1998].}

\reflbl\forrjanco{
	Forrester, P.J., and Jancovici, B.,
		{\it The two-dimensional two-component plasma
	plus background on a sphere: exact results},
	J. Stat. Phys. {\bf 84}, pp. 337--357 [1996].}

\reflbl\ginibre{
	Ginibre, J., 
		{\it Statistical ensembles of complex, quaternion, 
	and real matrices},
	J. Math. Phys. {\bf 6}, pp. 440--449 [1964].}

\reflbl\hewittsavage{
	Hewitt, E., and Savage, L. J., 
		{\it Symmetric measures on Cartesian products}, 
        Trans. Amer. Math. Soc. {\bf 80}, pp. 470--501 [1955].}

\reflbl\hsu{
	Hsu, P.L., 
		{\it On the distribution of roots of certain 
	determinantal equations}, Ann. Eug. {\bf 9}, pp. 240--258 [1939].}

\reflbl\kiesslingCPAM{
	Kiessling,  M. K.-H., 
		{\it Statistical mechanics of classical particles
        with logarithmic interactions,}
        Commun. Pure Appl. Math. {\bf 47}, pp. 27--56 [1993].}

\reflbl\kusnersullivan{
	Kusner, R.B., and Sullivan, J.M., 
		{\it M\"obius energies for knots and links, surfaces and 
	submanifolds}, Preprint [1995].}

\reflbl\kusuokatamura{
	Kusuoka, S., and Tamura, Y., 
		{\it Gibbs measures for mean field potentials}, 	
J. Fac. Sci. Univ. Tokyo, Sec. IA, Math. {\bf 31}, pp. 223--245 [1984].}
	
\reflbl\messerspohn{
	Messer, J., and Spohn, H., 
		{\it Statistical mechanics of the isothermal 
	Lane-Emden equation}, 
        J. Stat. Phys. {\bf 29}, pp. 561--578 [1982].}

\reflbl\mehtasriva{
	Mehta, M. L., and Srivastava, P.K., 
		{\it Correlation functions for eigenvalues of real quaternian
	matrices}, J. Math. Phys. {\bf 7}, pp. 341--344 [1966].}

\reflbl\mehtaBOOK{
	Mehta, M. L., 
		{\it Random Matrices}, 2nd ed., Acad. Press, New York [1991].}

\reflbl\johansson{
	K. Johansson, {\it On the fluctuations of eigenvalues of random 
	hermitian matrices}, Duke Math. J., {\bf 91}, pp.151--204 [1998].}

\reflbl\pasturshch{
	Pastur, L., and Shcherbina, M.,
		{\it Universality of the local eigenvalue statistics for a
	class of unitary invariant random matrix ensembles,}
	J. Stat. Phys. {\bf 86}, pp. 109--147 [1997].}

\reflbl\porterED{
	Porter, C. E. (Edit.), 
		{\it Statistical theories of spectra: fluctuations,} 
	New York, Academic Press, [1965].}

\reflbl\ruelleBOOK{
	Ruelle, D., 
		{\it Statistical Mechanics: Rigorous Results},
	Addison Wesley [1989].}

\reflbl\safftotikBOOK{
	Saff, E.B., and Totik, V.,
		{\it Logarithmic potentials with external fields},
	Grundl. d. Math. Wiss. vol. {\bf 316}, Springer [1997].}

\reflbl\sinaisosh{
	Sinai, Ya., and  Soshnikov, A.,  
		{\it Central limit theorem for traces of large random
	symmetric matrices with independent matrix elements}, 
	Preprint, Princeton Univ. [1997].}

\reflbl\spohnBOOK{
	Spohn, H., 
		{\it Large Scale Dynamics of Interacting Particles}, Texts
	and Monographs in Physics, Springer [1991]. }

\reflbl\jjthomson{
	Thomson, J.J., 
	Philos. Mag. {\bf 7}, p.237 [1904].}

\reflbl\wignerB{
	Wigner, E., 
		{\it Characteristic vectors of bordered matrices
	with infinite dimension}, 
	{\it Part I}, 	Ann. Math. {\bf 62}, pp. 548--564 [1955]; 
	{\it Part II}, 	ibid. {\bf 65}, pp. 203--207 [1957].}

\reflbl\wignerC{
	Wigner, E., 
		{\it On the distribution of the roots of certain 
	symmetric matrices},
	Ann. Math. {\bf 67}, pp. 325--327 [1958].}

\reflbl\wignerD{
	Wigner, E., 
		{\it Statistical properties of real symmetric matrices with 
	many dimensions,} 
	4th Can. Math. Congress (Banff 1957), Proc. pp. 174--184, 
	Univ. Toronto Press, [1959].}

\reflbl\wilson{
	Wilson, K. G., 
		{\it Proof of a conjecture by Dyson},
	J. Math. Phys. {\bf 3}, pp. 1040--1043 [1962].}
	
\endbib

 \centerline{\chhdsize A Note on the Eigenvalue Density of
			 Random Matrices}

\bigskip
\bigskip
\centerline{MICHAEL K.-H. KIESSLING$^1$ and HERBERT SPOHN$^{1,2}$}
\bigskip
\centerline{\it $^1$ Department of Mathematics,}
\centerline{\it Rutgers University,}
\centerline{\it 110 Frelinghuysen Rd., Piscataway, N.J. 08854}

\bigskip
\centerline{\it $^2$ On leave from: Theoretische Physik,}
\centerline{\it Ludwig-Maximilians-Universit\"at,} 
\centerline{\it Theresienstr. 37, D-80333 M\"unchen, FRG}

\bigskip
\centerline{ \it Electronic mail addresses: }
\medskip
\centerline{ Kiessling: \it miki@math.rutgers.edu }

\centerline{ Spohn: \it spohn@mathematik.tu-muenchen.de}

\bigskip\bigskip
\centerline{ABSTRACT}\smallskip
The distribution of eigenvalues of $N \times N$ random matrices in the 
limit $N\to \infty$ is the solution to a variational principle that 
determines the ground state energy of a  confined fluid of classical 
unit charges. This fact is 
a consequence of a  more general theorem, proven here, 
in the statistical mechanics of unstable interactions. Our result 
establishes the eigenvalue density of some ensembles of 
random matrices which were not covered by previous theorems.


\vskip 5truecm

\bigskip
\centerline{Original: March 11, 1998; revised: May 29, 1998}

\bigskip
\centerline{To appear in Comm. Math. Phys.}

\bigskip\bigskip\bigskip

\hrule
\medskip\noindent
Typeset in $\TeX$ by the authors. 
\bigskip\noindent
$\msimp{\copyright}{c}\,$ (1998) 
The authors. Reproduction of this article
for non-commercial purposes by any means is permitted.

\vfill\eject
 \chno=1

\baselineskip=18pt

\noindent
{\bf I. INTRODUCTION} 

\noindent
Since the pioneering work of Wigner [\wignerB,\wignerC,\wignerD], 
there has been a considerable 
effort to understand the statistics of eigenvalues of $N \times N$
random matrices. The problem has three scales: 
(i) the density of eigenvalues which converges to a deterministic limit 
	as $N\to \infty$;
(ii) the fluctuations of order one around this deterministic 
	density;
(iii) the local statistics on the scale of the typical distance between 
	eigenvalues. 
Item (i) depends on the particular matrix ensemble while (ii) and (iii) 
are ``universal,'' in the sense that they depend only on some overall 
matrix characteristics (e.g. the matrices being real and symmetric). 
The ``classical'' results are reviewed in [\mehtaBOOK]; 
see also [\porterED] for a collection of early work. Recently, 
in the context of the double scaling limit of 2D quantum 
gravity [\difrancesco],  (ii) and (iii) have been studied at the edge 
of the support of the density of states where novel universality 
classes occur. Among recent work on (ii) we also mention
[\johansson,\sinaisosh], and [\pasturshch,\bleherits,\deiftetal] 
regarding (iii). In our paper we will consider only the largest scale (i). 

For various $N\times N$ random matrix ensembles of the 
form
$$
m^{(N)}(d M) = {\bf Q}(N)^{-1}e^{-  \kappa N\, {\rm Tr}\, V(M)} dM\,  ,
\eqno\eqlbl\wishartmeas
$$
the joint probability distribution for the $N$ 
eigenvalues $\lambda_1,...,\lambda_N$ 
(which may be real or complex) is identical to the (configurational) 
canonical ensemble at inverse temperature $\beta$ of $N$ unit point charges 
at positions $\lambda_1,...,\lambda_N\in\Lambda\subset\RR^2$. The region
$\Lambda$ can be all of $\RR^2$, the unit disk ${\rm B}_1 \subset \RR^2$, 
the unit circle ${\SS}^1$, the entire real line $\RR$, or some other set, 
depending on the type of random matrices. 
This joint probability distribution has the general form
$$
d\mu^{(N)} = Q^{(N)}(\beta)^{-1}
 \exp\bigl(- \beta H^{(N)}\bigr) d\lambda_1\cdots d\lambda_N
\eqno\eqlbl\jEVmeas
$$
on $\Lambda^N$, where $d\lambda_k$ is the uniform measure on $\Lambda$ and 
$Q^{(N)}(\beta)$ the normalizing partition function. 
The classical Hamiltonian, $H^{(N)}$, is of the form
$$
H^{(N)}(\lambda_1,...,\lambda_N) =
\sum_{1\leq j<k\leq N} G(\lambda_j,\lambda_k)  +
\sum_{1\leq k\leq N}  F(\lambda_k) + N V(\lambda_k) \, ,
 \eqno\eqlbl\hamiltA
$$
where $G(\lambda_j,\lambda_k)= G(\lambda_k,\lambda_j)$
is ($2\pi\times$) a Green's function for $-\Delta$ in two dimensions,
$F(\lambda) =\lim_{\eta\to\lambda}(G(\lambda ,\eta) +\ln|\lambda -\eta|)$
is the regular part of $G$,  and $V(\lambda)$ given in (\wishartmeas).

Let us list a few examples. The joint eigenvalue distribution, $\mu^{(N)}$,
of the Gaussian ensembles (i.e., $V(M) = M^{\dagger}M$ in (\wishartmeas)) 
for the real symmetric [\hsu], the complex Hermitian [\dysonB], the general 
complex [\ginibre], and the Hermitian self-dual quaternionic [\dysonB] 
$N\times N$ random matrices takes the form 
$$
d\mu^{(N)} = {1\over Q^{(N)}(\beta)}
\prod_{1\leq i<j\leq N} |\lambda_i - \lambda_j|^\beta\prod_{1\leq k\leq N} 
e^{-\beta N |\lambda_k|^2} d\lambda_k\, ;
\eqno\eqlbl\specmeasHERM
$$
see also [\mehtaBOOK]. Clearly, this corresponds to (\jEVmeas), (\hamiltA), 
where $G(\lambda_1,\lambda_2) = -\ln |\lambda_1 -\lambda_2|$ is the 
free space Green's function (whence $F\equiv 0$)
and $V(\lambda) = |\lambda|^2$ 
a quadratic potential.  The eigenvalues for real symmetric, complex 
Hermitian, and Hermitian self-dual quaternionic matrices are  real.
Therefore, the charges  are confined to $\RR$, i.e. $\Lambda = \RR$. 
The parameters have the values $\beta =1,2,4$ and $\kappa =1,2,2$, 
respectively. For each of the associated  ensembles of unitary matrices the
charges are confined to the unit circle, i.e. $\Lambda = \SS^{1}$. 
Then $\beta$ is unmodified but $\kappa =0$ [\dysonA].  
For general complex random matrices, the eigenvalues 
are complex, corresponding to  unconfined charges, i.e. 
$\Lambda =\RR^2 =\CC$, and  $\beta =2$, $\kappa =2$ [\ginibre].
The joint eigenvalue distribution for the general real quaternionic 
$N\times N$ matrices takes the more complicated form [\ginibre]
$$
d\mu^{(N)} =  {1\over Q^{(N)}(\beta)} \prod_{1\leq i<j\leq N}  
|\lambda_i - \lambda_j|^\beta |\lambda_i - \lambda_j^*|^\beta
\prod_{1\leq k\leq N} |\lambda_k - \lambda_k^*|^\beta
e^{- \beta N |\lambda_k|^2}  d\lambda_k 
\eqno\eqlbl\specmeasQ 
$$
with $\beta = 2$, $\kappa =1$, $\Lambda = \RR^2 =\CC$.
Although apparently not noticed previously, 
(\specmeasQ) can be interpreted as a configurational canonical 
Coulomb ensemble, with a Hamiltonian (\hamiltA)
in which now $G(\lambda_1,\lambda_2) = -\ln |\lambda_1 -\lambda_2|
-\ln |\lambda_1 -\lambda_2^*|$ is the Green's function of 
$-\Delta$ for the upper half space $\RR\times\RR^+ = \CC^+$ 
equipped with a perfectly dia-electric condition at its boundary 
$\partial\CC^+$ (the real axis) and asymptotic free conditions 
at infinity, extended symmetrically to $\RR^2$, having a regular
part given by $F(\lambda) = - \ln (2 |\Im(\lambda)|)$. Moreover,
$V(\lambda) = |\lambda|^2$. This Hamiltonian describes
$N$ Coulomb point charges interacting via the free space Green's 
function $\, -\! \ln |\lambda - \lambda^\pr|$ amongst each other
and also with $N$ identical image charges  with respect to the line 
$\Im (\lambda) =0$. Since the interaction of a charge with its own 
image contributes only an amount $F/2$ to the Hamiltonian, 
$F/2 +NV$ is now to be counted as the external potential.  

For symmetric random matrices, properties of the eigenvalue statistics 
have been computed in great detail for arbitrary $N$, using explicit 
expansion techniques [\mehtaBOOK, \mehtasriva, \ginibre], group 
theoretical methods [\dysonA], the method of orthogonal polynomials 
[\mehtaBOOK,\pasturshch,\bleherits,\deiftetal], as well as some more 
recent developments in soliton theory and two-dimensional quantum gravity 
[\adlerA,\adlerB].
The beautiful connection to two-dimensional Coulomb systems suggests 
to use the general methods of statistical mechanics when the above 
algebraic methods fail. Even in the exactly solvable situations the 
statistical mechanics approach may provide us rather readily
with certain relevant asymptotic ($N\to\infty$) results which to 
extract from the exact finite $N$ solutions would require quite tedious and 
lengthy computations. Furthermore, as Dyson has pointed out [\dysonA],
in the framework of statistical mechanics the limit $N\to\infty$ makes sense
for arbitrary $\beta$ and not only for the 
discrete values of $\beta = 1,2,4$. Thereby one achieves
a ``thermodynamic view point''
which yields valuable new insights into random matrices. Early 
results in this direction are in [\dysonA,\wilson], and more 
recent ones in [\costinjll,\foglershklovskii,\forrester]. 

The prime example of the exploitation of the Coulomb analogy 
is Wigner's [\wignerD] electrostatic derivation of his semi-circle law
$$
\rho(\lambda)=
\left\{
	{(2/\pi)(1 - \lambda^2)^{1/2}\, ;\quad |\lambda| <    1 \atop 
	\qquad\qquad 0\qquad \quad ;\quad   |\lambda| \geq 1} 
\right.  \ 
\eqno\eqlbl\sclaw
$$
for the eigenvalue density $\rho(\lambda)$ in the limit $N = \infty$ 
of the Gaussian ensemble (\wishartmeas) in 
the case of real symmetric random matrices. Wigner [\wignerD] argued, 
heuristically, that when $N\to\infty$, the eigenvalue density 
for (\specmeasHERM) can be obtained from a variational principle for 
a continuum charge density $\rho(\lambda)$ of total charge 1, 
restricted to the real line, that satisfies the requirement
of mechanical force balance between its own electrostatic force field 
and the applied force field $ - \partial_\lambda |\lambda|^2 =-2\lambda$. 
Previously [\wignerB] he had proved, by the method  
of moments, that (\sclaw) holds true for a Bernoulli ensemble 
of bordered random sign matrices, which suggested that (\sclaw)
was the limiting law under more general circumstances [\wignerD].  
In [\wignerC,\wignerD] he announced that (\sclaw) can indeed be 
proved to hold for a wider class of ensembles under a mild set of 
conditions, including the Gaussian real symmetric ensembles. 

Interestingly in itself,
entropy plays no role in Wigner's variational principle, 
which is concerned only with the ground state energy  of the classical 
continuum Coulomb fluid. To get an intuitive idea how this can arise from  
the canonical measure (\specmeasHERM) with {\it fixed} $\beta $ 
($<\infty )$, we rewrite (\specmeasHERM) as 
$$
d\mu^{(N)} = {1\over Q^{(N)}(\beta)}\exp\Biggl(\beta N\Biggl[{1\over N}
\sum_{1\leq i<j\leq N} \ln |\lambda_i - \lambda_j| - 
\sum_{k=1}^N  |\lambda_k|^2 \Biggr]\Biggr)
\prod_{1\leq k\leq N}  d\lambda_k\, .
\eqno\eqlbl\MFmeas
$$
Apparently, the limit $N\to\infty$ for (\MFmeas) is a 
{\it simultaneous} thermodynamic {\it and} zero-temperature limit 
for an unstable Hamiltonian with mean-field scaling. If in (\MFmeas) 
we replace $\beta N$ by  $\beta N_0$, with $N_0$ fixed, and then
let $N \to \infty$, we obtain the variational principle of
[\clmpCMPcan,\kiesslingCPAM] for a continuum free energy 
functional at inverse mean-field temperature $\beta_{MF} = \beta N_0$, 
see also [\messerspohn].
Letting $N_0 \to\infty$ subsequently, the entropy  contribution to the 
free energy drops out, giving formally Wigner's  variational principle. 
With a leap of faith one may thus expect that the limit $N\to\infty$
in (\MFmeas) will give the same result directly.

Recently, Boutet de Monvel, Pastur, and Shcherbina [\boutetetal] 
have studied the limit $N\to \infty$ 
of the joint eigenvalue distribution
of  real symmetric random matrix ensembles (\wishartmeas) 
for a large class of $V(M)$, satisfying certain regularity conditions. 
They prove that in the limit $N\to \infty$ the $n$-th marginal measure 
of (\specmeasHERM), with $\lambda^2$ replaced by $V(\lambda)$, 
factors into an 
$n$-fold tensor product of identical one-particle measures 
whose density is precisely that of the two-dimensional Coulomb fluid
restricted to a line, satisfying the equations of 
electrostatic equilibrium in the applied potential field $V$. 
The proof in [\boutetetal] is based 
on the classical Stieltjes transform. It is carried out
explicitly for $\beta =1$, but the method  covers all
$\beta > 0$ (in particular, including $\beta =2,4$),  
as well as certain H\"older continuous many-body interactions. 

We here generalize this result in [\boutetetal] 
to a wider class of interactions and an arbitrary (finite) number of 
space dimension. Our result covers also the limit $N\to\infty$ 
of the eigenvalue distribution of Ginibre's complex and real 
quaternionic Gaussian random matrix ensembles [\ginibre].
However, our method is very different from that in [\boutetetal].  
Instead of using the Stieltjes transform, we adapt the strategy of 
Messer and Spohn [\messerspohn], Kiessling [\kiesslingCPAM] 
and Caglioti et al.[\clmpCMPcan] to the combined mean-field 
and zero temperature limit. Interestingly enough, we can work
without the detailed control of [\boutetetal] on the finite-$N$ 
marginal densities, and in this sense our proof of the 
variational principle is also considerably shorter and simpler 
than the proof in [\boutetetal].

\vfill\eject

\smallskip
\noindent
{\bf II. MAIN RESULT} \chno=2 \equno=0

We now prepare the statements of our main result. Let
$\Lambda\subset\RR^{\rm d}$ be closed and connected, and let $dx$ 
denote uniform measure on $\Lambda$. Note that $\Lambda$ may be all 
of $\RR^{\rm d}$, or it may be a lower dimensional 
manifold, e.g. the sphere $\SS^{\rm d-1}$. 
We denote by $P(\Lambda)$ the probability measures on $\Lambda$, 
and by $P^s(\Lambda^\NN)$ the permutation symmetric probability 
measures on the infinite Cartesian product ${\Lambda}^\NN$. 	
We recall the decomposition theorem of de Finetti [\finetti] 
and Dynkin [\dynkin] (see also [\hewittsavage,\ellisBOOK]), 
which states that $\mu\in P^s(\Lambda^\NN)$ 
is uniquely presentable as a linear convex 
superposition of product measures, i.e., for each
$\mu \in P^s(\Lambda^\NN)$ there exists a unique 
probability measure $\nu(d\varrho|\mu)$ on $P(\Lambda)$, such
that for each $n\in \NN$,
$$
\mu_n (d^{n}x) =
	\int_{P(\Lambda)} \nu(d\varrho|\mu) 
		\varrho^{\otimes n}(d^{n}x)  \, ,
\eqno\eqlbl\finettimeas
$$
where $\mu_n$ denotes the $n$-th marginal measure of $\mu$. 
This is also the extremal decomposition for the convex  
set $P^s(\Lambda^\NN)$, see [\hewittsavage]. 

To establish the limit $N\to\infty$ for measures of the form
(\jEVmeas), (\hamiltA) only some general properties of the
interactions enter. Also, the particular value of 
$\beta$ plays no role, and it is convenient to absorb it into
the Hamiltonian. Therefore, we study the sequence of
probability measures 
$$
\mu^{(N)}\left(d^{N}x\right) = {1\over Q^{(N)}}
	\exp\left(-  H^{(N)}({\bf x}_1,...,{\bf x}_N)\right) 
	\prod_{1\leq k\leq N} dx_k
\eqno\eqlbl\canonmeas
$$
for Hamiltonians of the form 
$$
 H^{(N)}({\bf x}_1,...,{\bf x}_N)  = 
	\sum_{1\leq i<j \leq N} w({\bf x}_i, {\bf x}_j )
	+  \sum_{1\leq i \leq N}u({\bf x}_i) + N v({\bf x}_i)\, .
\eqno\eqlbl\hamiltB
$$
The pair interaction $w$ and one-particle potentials $u$, $v$ 
satisfy the following conditions.
\medskip\noindent
Condition on $w(\bx,\by)$: 
$$
\leqno(C1)\qquad\qquad {\it Symmetry:}
		\ w(\bx,\by)=w(\by,\bx) 
$$
\medskip\noindent
Condition on $U(\bx) = u(\bx) + v(\bx)$:
$$
\leqno(C2) \qquad\qquad {\it Integrability:}
	\ e^{-  U(\bx)} \in L^1(\Lambda,dx)\ 
$$
\medskip\noindent
Conditions on $W(\bx,\by)=w({\bf x},{\bf y})+v({\bf x})+ v({\bf y})$:
$$
\leqno(C3)\quad\qquad {\it  Lower\ semicontinuity:}
	\ W\ {\rm is\ l.s.c.\ on\ }  {\Lambda} \times {\Lambda}
$$
$$
\leqno(C4)\quad\qquad {\it Integrability:}
	\ W(\bx,\by)\in L^1\(\Lambda^2,\, 
e^{- U(\bx)}dx \otimes e^{- U(\by)} dy\)\ 
$$
In case of an unbounded $\Lambda$ we also need a growth condition
at infinity. Let $W_-(\bx) \equiv \min_{\by} W(\bx,\by)$. 
$$
\leqno(C5) \quad\qquad  {\it Confinement:} 
	\ {\displaystyle \lim_{|\bx |\to \infty}} W_-(\bx) = \infty \, ,\
		{\rm uniformly\ in\ }\bx 
$$

\smallskip
\noindent
{\bf THEOREM:} {\it Let $\Lambda\subset\RR^{\rm d}$ be closed and connected,
and let $w$, $v$ and $u$ 
satisfy the conditions $(C1)-(C5)$.
Consider (\canonmeas) as extended to a probability on $\Lambda^\NN$. 
Then there exists a $\mu \in P^s(\Lambda^\NN)$ such that,
after extraction of a  subsequence $\mu^{(N^\pr)}$, 
$$
\lim_{N^\pr\to\infty} \mu^{(N^\pr)} = \mu \, .
\eqno\eqlbl\measurelim
$$
For each limit point $\mu$, the decomposition measure  
$\nu(d\varrho|\mu)$ is concentrated on the subset of 
$P(\Lambda)$ which consists of the probability measures 
$\varrho$ that minimize the functional
$$
\cE(\varrho) = {1\over 2}\varrho^{\otimes 2}(W) 
\eqno\eqlbl\energyVP
$$ 
over $P(\Lambda)$.}

In general we have little information on the decomposition measure 
$\nu(d\varrho|\mu)$. More is known for regular mean-field Hamiltonians, 
see [\kusuokatamura]. However, if it can be shown, as is the case for 
many random matrix ensembles, that (\energyVP) has a unique minimizer, say 
$\varrho_{0}$, then in (\measurelim) we have in fact convergence, and 
the limit is of the form
$$
\mu_n  = \varrho_{0}^{\otimes n}.  
\eqno\eqlbl\fac
$$

As discussed in [\spohnBOOK], the factorization property (\fac) is 
equivalent to a law of large numbers. Consider the eigenvalues 
averaged over some continuous test function $f$, 
$$
\bigl\langle f \bigr\rangle_N 	
\equiv {1 \over N} \sum_{j=1}^{N}f(\lambda_{j})\, .
\eqno\eqlbl\average
$$
Then (\fac) implies that, for all such $f$,
$$
\lim _{ N \to \infty}\bigl\langle f \bigr\rangle_N 	
= \int_\Lambda \varrho_{0}(d\lambda ) f(\lambda )
\eqno\eqlbl\lln
$$
in probability. We summarize as
\bigskip
\noindent
{\bf COROLLARY:} {\it If (\energyVP) has a unique minimizer, $\varrho_{0}$, 
then the weak law of large numbers (\lln) holds for all $f\in C^{0}(\Lambda)$.}

\medskip

Applications of our theorem to random matrix ensembles
are presented in the concluding section.

\bigskip

\noindent
{\bf III. PROOF OF THE THEOREM} \chno=3 \equno=0

The proof looks somewhat technical because we work under fairly minimal
assumptions on $W$. But, in essence, we only have to show, through sharp
upper and lower bounds, that the pair-specific free energy converges 
to the minimum continuum energy. The remainder of the theorem 
follows from the permutation invariance of the measures. 

We define the absolutely continuous (w.r.t. $dx$)
a-priori probability measure on $\Lambda$,
$$
\mu_0(dx) =  Z_0^{-1} e^{- U(\bx)} dx \, .
\eqno\eqlbl\apriorimeas
$$
For each $\varrho^{(N)} \in P(\Lambda^N)$  its entropy w.r.t.
$\mu_0^{\otimes N}$ is defined by
$$
\cS^{(N)}\left(\varrho^{(N)}\right) =
 - \int_{\Lambda^N} 
\ln \left({d\varrho^{(N)} \over d\mu_0^{\otimes N}} \right)
 \varrho^{(N)}(d^Nx)
\eqno\eqlbl\relentropy
$$
if $\varrho^{(N)}$ is absolutely continuous w.r.t. a-priori measure 
$\mu_0^{\otimes N}$, and provided the integral in (\relentropy) exists. 
In all other cases, $\cS^{(N)}\left(\varrho^{(N)}\right) = -\infty$.

\smallskip\noindent
{\bf LEMMA 1:} {\it The relative entropy (\relentropy) is non-positive,
$$
\cS^{(N)}\bigl(\varrho^{(N)}\bigr) \leq 0\, .
\eqno\eqlbl\entropybound
$$}
{\bf Proof of Lemma 1:} 

A standard convexity argument, see [\ellisBOOK]. \eqed
\medskip

We introduce the symmetric Hamiltonian 
$$
K^{(N)}({\bf x}_1,...,{\bf x}_N)
 = {1\over 2} \sum_{1\leq j\neq k \leq N}  W(\bx_j,\bx_k)\, .
\eqno\eqlbl\Khamilton
$$
For $\varrho^{(N)} \in P(\Lambda^N)$, ($\beta \times$)
its Helmholtz free energy (or just free energy) is now defined by
$$
\cF^{(N)}\bigl(\varrho^{(N)}\bigr) = 
 \varrho^{(N)}\bigl( K^{(N)}\bigr) - \cS^{(N)}\bigl(\varrho^{(N)}\bigr) 
\eqno\eqlbl\FE
$$
if the right side exists, and by $\cF^{(N)}\left(\varrho^{(N)}\right) =
+\infty$ in all other cases. 

\smallskip\noindent
{\bf LEMMA 2:} {\it The free energy (\FE) takes its unique minimum at
the probability measure 
$$
\mu^{(N)}(d^Nx) =  {1 \over Z^{(N)}}
\exp\left(- K^{(N)}({\bf x}_1,...,{\bf x}_N)\right) 
\mu_0^{\otimes N} (d^Nx)\, ,
\eqno\eqlbl\meas
$$
where ${Z}^{(N)} = Q^{(N)}/Z_0^N$. Thus,
$$
\min_{\varrho^{(N)}\in P(\Lambda^N)}
\cF^{(N)}\bigl(\varrho^{(N)}\bigr) = 
\cF^{(N)}\bigl(\mu^{(N)}\bigr) = - \ln {Z}^{(N)}\, .
\eqno\eqlbl\vpFE
$$
} 

\medskip\noindent
{\bf Proof of Lemma 2:} 

The variational principle is verified by 
a standard convexity estimate [\ellisBOOK,\ruelleBOOK], 
which shows that $\cF^{(N)}\bigl(\varrho^{(N)}
\bigr) -\cF^{(N)}\bigl(\mu^{(N)}\bigr) \geq 0$, 
with equality holding if and only if $\varrho^{(N)} = \mu^{(N)}$. 
The second identity in (\vpFE) is verified by explicit calculation. 
\eqed

Notice that the canonical probability measure (\meas) is
just (\canonmeas) rewritten in terms of $K^{(N)}$ and $\mu_0$. 

\smallskip\noindent
{\bf LEMMA 3:} {\it The functional $\cE(\varrho)$ has a finite
minimum over $P(\Lambda)$.}

\medskip\noindent
{\bf Proof of Lemma 3:} 

For bounded $\Lambda$ the lower semicontinuity $(C3)$ establishes
that $\cE(\varrho)$ is a lower semicontinuous functional for the 
topology of measures in  $P(\Lambda)$. 
For an unbounded region, the same conclusion holds because
of $(C3)$ and $(C5)$. Now the claim follows from standard facts about
lower semicontinuous functionals [\ruelleBOOK]. \eqed

\medskip

In the following, let $\varrho_0$ denote a minimizing probability measure 
for (\energyVP), and let $E_0 = \cE(\varrho_0)$. 

\smallskip\noindent
{\bf LEMMA 4:} {\it The pair specific free energy is bounded above by
$$
\limsup_{N\to \infty} \(- N^{-2} \ln {Z}^{(N)} \)
	\leq  E_0 \, .
\eqno\eqlbl\upbound
$$}

We will prove Lemma 4 in two blocks, distinguishing minimizers $\varrho_0$ 
with finite entropy from those with negative infinite entropy. For instance, 
Wigner's semicircle density (\sclaw) has finite entropy, but other ensembles 
can have a singular measure as eigenvalue ``density.'' 


\medskip\noindent
{\bf Proof of Lemma 4 }

\noindent
Assume first that the entropy of $\varrho_0$ is finite, 
i.e. (recalling  Lemma 1), 
$$
\cS^{(1)}\(\varrho_0\)  =  S_0 \in (-\infty\, ,0]\, .
\eqno\eqlbl\Snullvalue
$$
Notice that the non-positive constant $S_0$ is $N$-independent.     
We then can use Lemma 2 to estimate for all $N$ that 
$$
- N^{-2} \ln{Z}^{(N)}
\leq  N^{-2} \cF^{(N)}(\varrho_0^{\otimes N})
=\(1-  N^{-1}\) \cE(\varrho_0)- N^{-1} \cS^{(1)}\(\varrho_0\)\, .
\eqno\eqlbl\upest 
$$
The identity in (\upest) follows from $(C1)$. By (\Snullvalue), 
$\cS^{(1)}\(\varrho_0\) = S_0$, and $S_0$ is finite and
$N$-independent, and by Lemma 3, $\cE(\varrho_0) = E_0$ is finite and
$N$-independent. Thus (\upbound) in the finite entropy case
 follows by taking $N\to\infty$ in (\upest). 

Assume now that (\Snullvalue) is false, so that
$\cS^{(1)}(\varrho_0) = -\infty$. In that case, (\upest) becomes useless. 
Now, since the $C_0^\infty$ functions are dense in $P(\Lambda)$, the
obvious way out is to modify  the above argument and 
to work with a regular approximation to $\varrho_0$.  
However, since $\cE(\varrho)$ is only lower semicontinuous, we 
have to employ also a continuous approximation to $\cE(\varrho)$. 

By $(C3)$ for bounded $\Lambda$, 
and by $(C3)$, $(C5)$ in case of an unbounded 
$\Lambda$, $W$ is the pointwise upper limit of a continuous increasing
map $\gamma\mapsto W_\gamma \in C^0({\Lambda}\times{\Lambda})$
that is uniformly bounded below, see [\ruelleBOOK]. By $(C1)$, we can 
assume that $W_\gamma(\bx,\by) = W_\gamma(\by,\bx)$. In the following,
let $K_\gamma^{(N)}$ be defined by (\Khamilton) with $W$ replaced 
by $W_\gamma$, and let $\mu_\gamma^{(N)}\(d^Nx\)$ and
$Z_\gamma^{(N)}$, 
respectively $\cF_{\gamma}^{(N)}\left(\varrho^{(N)}\right)$,
be defined by (\meas), respectively (\FE), 
with $K^{(N)}$ replaced by $K_\gamma^{(N)}$. 

Since $W$ satisfies $(C3)$, and $W_\gamma$ is of class $C^0$, and
$W_\gamma \nearrow W$ pointwise, for each positive $\eps \ll 1$ 
we can find a $\gamma_\eps$ such that, simultaneously,
$$
E_0 - {1\over 2} \varrho_0^{\otimes 2} (W_\gamma) < {1\over 3} \eps 
\eqno\eqlbl\Eappr
$$
and 
$$
\limsup_{N\to\infty} \(-N^{-2}\ln {Z}^{(N)} \) \leq 
\limsup_{N\to \infty} \(-N^{-2}\ln {Z}_\gamma^{(N)}\)+
{1\over 3}\eps
\eqno\eqlbl\Zappr
$$
whenever $\gamma \geq \gamma_\eps$.  Moreover, since $W_\gamma$
is  bounded below on ${\Lambda}\times {\Lambda}$ uniformly 
in $\gamma$, and $W_\gamma \in C^0({\Lambda}\times {\Lambda})$,
for each $\gamma_\eps$ we can find a measure 
$\varrho_\eps^{\phantom{x}}\in P(\Lambda)$ that is 
equivalent to a positive function  of class $C_0^\infty(\Lambda)$, such that 
$$
\left| {1 \over 2} \varrho^{\otimes 2}_\eps(W_{\gamma_\eps}) - 
{1 \over 2} \varrho^{\otimes 2}_0(W_{\gamma_\eps}) \right| < 
{1 \over 3} \eps\, .
\eqno\eqlbl\nEappr
$$

On the other hand, given any $\gamma$ and any 
$\varrho_\delta^{\phantom{x}}\in P(\Lambda)$ of class $C_0^\infty$, 
we have the estimate
$$
\eqalignno{
 - \ln {Z}^{(N)}_\gamma
 &  =   \cF_{\gamma}^{(N)}\bigl(\mu_\gamma^{(N)}\bigr) 
	=  \min_{\varrho^{(N)}\in P(\Lambda^N)} 
		\cF_{\gamma}^{(N)}\bigl(\varrho^{(N)}\bigr) &\cr
&\leq\, \cF_{\gamma}^{(N)}\left(\varrho_\delta^{\otimes N}\right) 
   = N (N-1) {1\over 2}  \varrho_\delta^{\otimes 2}( W_\gamma) 
		+ N \cS^{(1)}(\varrho_\delta^{\phantom{x}})\, ,
&\eqlbl\ESTIM \cr}
$$
where the first line is the analog of Lemma 2, the inequality obvious, and
the last identity an explicit computation, using the symmetry of the 
$W_\gamma$. In particular, for any given $\eps$ we can choose 
$\gamma = \gamma_\eps$ and 
$\varrho_\delta^{\phantom{x}} = \varrho_\eps^{\phantom{x}}$ in
(\ESTIM), then multiply (\ESTIM) by $N^{-2}$ and take the limsup. 
Since $\varrho_\eps^{\phantom{x}}$ is of class $C_0^\infty$, we have 
$|\cS^{(1)}(\varrho_\eps^{\phantom{x}})| = C(\eps) < \infty$, independent 
of $N$, whence $N^{-1}\cS^{(1)}(\varrho_\eps^{\phantom{x}})\to 0$. Next
we use (\Eappr), (\Zappr), (\nEappr), and the triangle inequality, and
conclude
$$
\limsup_{N\to \infty} 
\(- N^{-2} \ln {Z}^{(N)} \) 
\leq  {1\over 2} \varrho_\eps^{\otimes 2}(W_{\gamma_\eps}) 
	+ {1\over 3} \eps \leq   E_0 + \eps \, ,
\eqno\eqlbl\updeltaepsbound
$$
for arbitrarily small $\eps$. This proves (\upbound)
for the infinite entropy case.

The proof of Lemma 4 is complete. \eqed


\smallskip\noindent
{\bf LEMMA 5:} 
{\it The sequence $N\mapsto \mu^{(N)}$ given by (\meas) 
is compact for  bounded $\Lambda$, and tight for unbounded $\Lambda$.}
\medskip

 \vfill\eject

\noindent
{\bf Proof of Lemma 5:} 

If $\Lambda$ is bounded then it is also compact, for 
$\Lambda\subset \RR^{\rm d}$ is closed, and in that case $P(\Lambda)$ 
is compact for the topology of measures. By Tychonov's theorem, 
$P^s(\Lambda^\NN)$ is now compact in the product topology. Hence, 
for bounded $\Lambda$, the sequence (\meas) is compact. 

If $\Lambda$ is unbounded, we have to estimate the 
contribution from outside of ${\rm B}_R^{\ n}$ to the mass 
of the $n$-th marginal $\mu_n^{(N)}$ of (\meas), 
when $N\to\infty$. Recall that a sequence of
probability measures $\mu_n^{(N)}$ is tight if for each $\eps\ll 1$ there
exists a $R=R(\eps)$ such that $\mu_n^{(N)}({\rm B}_R^{\ n}) > 1-\eps$,
independent of $N$, see [\durrettBOOK]. Since our marginal measures are
compatible 
(i.e., $\mu_n^{(N)}(d^nx) = \mu_{m}^{(N)}(d^nx\otimes\Lambda^{m-n})$ 
for $m>n$) and permutation symmetric by $(C1)$,
it suffices to prove tightness for any particular $n$. We pick $n=2$. 

We notice that by $(C3)$ and $(C5)$, 
$$
 \min_{(\bx,\by)\in {\Lambda\times\Lambda}} W(\bx,\by) = W_0 > -\infty \, . 
\eqno\eqlbl\minW 
$$
Since (\meas) is invariant under the transformation $W \to W+ C$, 
we can even assume, without loss of generality, that $W_0 > 0$. 
We then have the following sandwich bounds,  independent of $N$, 
$$
0 < \mu_2^{(N)}(W) \leq \mu_0^{\otimes 2}(W)\, ,
\eqno\eqlbl\sandwEST
$$
with $\mu_0^{\otimes 2}(W)< \infty$, by $(C4)$.  The lower bound 
in (\sandwEST) is obvious, for $W_0 >0$. To prove the upper bound 
in (\sandwEST) we use the strategy of [\kiesslingCPAM]. We can 
replace $K^{(N)}$ by $\alpha K^{(N)}$ in (\meas), with 
$0\leq \alpha <\infty$, so that 
$- 2 N^{-2}(1-N^{-1})\ln {Z}^{(N)}=\Gamma_N(\alpha)$ is now a function 
of $\alpha$. Clearly, $\Gamma_N(0) =0$, and $W\geq 0$ implies 
$\Gamma_N(\alpha)\geq 0$ as well as $\Gamma_N^\prime(\alpha) \geq 0$, while
the Cauchy-Schwarz inequality implies $\Gamma_N^\ppr(\alpha) \leq 0$. 
Moreover, Jensen's inequality, applied w.r.t. $\mu_0^{\otimes N}$,  
and $(C4)$ imply 
$$
\Gamma_N(\alpha) \leq \alpha \mu_0^{\otimes 2}(W)\, .
\eqno\eqlbl\uppGest
$$ 
Obviously $\mu_0^{\otimes 2}(W)$ is $N$-independent. 
Thus, $\Gamma_N(\alpha)$ is a nonnegative, increasing, concave real function, 
bounded above by (\uppGest),  and mapping zero into itself.
A simple geometrical argument now reveals that the slope
of any tangent to $\Gamma_N(\alpha)$  never exceeds the slope of 
the ray  on the r.h.s. of (\uppGest), i.e.,
$\Gamma_N^\prime(\alpha) \leq \mu_0^{\otimes 2}(W)$. 
But $\Gamma_N^\prime(1) =  \mu_2^{(N)}(W)$, which proves 
the right inequality in (\sandwEST). 

Now pick $\eps \ll 1$ arbitrary. By $(C5)$ we can find a $R=R(\eps)$
such that 
$$
\inf_{(\bx,\by)\ \not\in \ {\rm B}_R^2} W(\bx,\by) \geq 
{1\over \eps}\mu_0^{\otimes 2}(W) \, .
\eqno\eqlbl\Uratio
$$
Let $\chi$ denote the characteristic function of the complement of
${\rm B}_R$ in $\Lambda$. We then have the chain of estimates
$$
\eqalignno{
 \mu_0^{\otimes 2}(W) 
& \geq \mu_2^{(N)}(W) \geq \mu_2^{(N)}\bigl(W \chi^{\otimes 2}\bigr) 
&\cr & 
\geq
	 \inf_{(\bx,\by)\ \not\in \ {\rm B}_R^2} W(\bx,\by) \, 
		 \mu_2^{(N)}\bigl( \chi^{\otimes 2}\bigr) 
 \geq {1\over \eps}\mu_0^{\otimes 2}(W) \(1 - \mu_2^{(N)}\bigl({\rm
B}_R^{\ 2} \bigr) \)\, .
&\eqlbl\chainEST \cr}
$$
Division of (\chainEST) by $\eps^{-1} \mu_0^{\otimes 2}(W)$ and 
a simple rewriting reveals that, independent of $N$, 
$$
\mu_2^{(N)}\bigl({\rm B}_R^{\ 2} \bigr)  \geq 1 - \eps\, ,
\eqno\eqlbl\tightness
$$
which was to be shown. The proof is complete. \eqed

 \smallskip
\noindent
{\bf LEMMA 6:} {\it The pair specific free energy is bounded below by
$$
\liminf_{N\to \infty} \( - N^{-2} \ln {Z}^{(N)}\)
	\geq  E_0 \, .
\eqno\eqlbl\lowbound
$$}

\noindent
{\bf Proof of Lemma 6:} 

By Lemma 1, $\cS^{(N)}(\mu^{(N)}) \leq 0$. Therefore,
$$
- \ln {Z}^{(N)}  \geq  \mu^{(N)}\bigl(  K^{(N)}\bigr)\, .
\eqno\eqlbl\lowestA
$$
By $(C1)$, 
$$
{1\over N^2}\mu^{(N)}\bigl( K^{(N)}\bigr)
	= \(1 -{1\over N}\) {1\over 2} \mu_2^{(N)}(W)\, .
\eqno\eqlbl\lowestB
$$
Now pick a converging subsequence of (\meas), 
$\mu^{(N^\pr)} \rightharpoonup \mu\in P^s(\Lambda^\NN)$. 
Such a  converging subsequence exists  by Lemma 5 and the  
Bolzano-Weierstrass theorem.  Then,  by $(C3)$, we have
$$
\liminf_{N^\prime\to\infty}  \mu_2^{(N^\prime)}(W) \geq  \mu_2(W) \, ,
\eqno\eqlbl\lowestC
$$
while $1- {N^\prime}^{-1} \to 1$ trivially. Thus,
$$
\liminf_{N\to \infty} \( - N^{-2} \ln \bigl(Z^{(N)}\bigr)\) 
	\geq    {1\over 2} \mu_2 (W)\, .
\eqno\eqlbl\lowbound
$$

Finally,  using the representation (\finettimeas), we see that
$$
{1\over 2} \mu_2(W) =
\int_{P(\Lambda)} \nu(d\varrho|\mu)\, \cE(\varrho) \geq \cE(\varrho_0)\, ,
\eqno\eqlbl\energylow
$$
and the proof of Lemma 6 is complete. \eqed

\smallskip
\noindent
{\bf Proof of the Theorem.}

By Lemma 4 and Lemma 6, 
$$
\lim_{N\to \infty} \bigl( - N^{-2} \ln {Z}^{(N)} \bigr) =   E_0 \, .
\eqno\eqlbl\FElim
$$
Recalling (\lowbound) and (\energylow), we see that (\FElim)
implies  
$$
\int_{P(\Lambda)} \nu(d\varrho|\mu)\, \cE(\varrho) = \cE(\varrho_0)\, 
\eqno\eqlbl\energydefinetti
$$
for every limit point $\mu$ of $\mu^{(N)}$. 
Equation (\energydefinetti) in turn implies that
the decomposition measure $\nu(d\varrho|\mu)$ is concentrated 
on the minimizers of $\cE(\varrho)$, for assume not, then 
$$
\int_{P(\Lambda)} \nu(d\varrho|\mu)\, \cE(\varrho) > \cE(\varrho_0) \, ,
$$
which contradicts (\energydefinetti). The proof of the Theorem is
complete.\eqed

We are now also in the position to vindicate our remark on
the existence of the limit in (\measurelim) in case
the minimizer $\varrho_0$ is unique. Indeed, in that case the
set of limit points of $\{ \mu^{(N)}, N=1,2,\ldots \}$ consists of 
the single measure.

\smallskip

\noindent
{\bf IV. APPLICATIONS } \chno=4 \equno=0

With the specifications in (\canonmeas), (\hamiltB) 
that $\bx = \lambda \in \Lambda\subset\RR^2$, 
$v(\lambda) = \beta V(\lambda)$, $u(\lambda) = \beta F(\lambda)$, 
and $w(\lambda,\eta) = \beta G(\lambda,\eta)$, 
where $G$ is a Green's function for $-\Delta$ in 2D and $F$ its regular 
part, our theorem characterizes the limit $N=\infty$ of (\jEVmeas), 
(\hamiltA), which for $\beta = 1,2,4$ is the joint eigenvalue 
distribution of various random matrix ensembles of the form (\wishartmeas). 
The decomposition measure of the limit is concentrated on the 
ground state(s) of the electrostatic energy functional 
$\veps(\varrho) = \beta^{-1} \cE(\varrho)$ of a charged 
continuum fluid with ``charge density'' $d\varrho/d\lambda$
(which may be a singular measure) of total charge 1, subject 
to an external potential $V$. Explicitly, the energy functional reads
$$
\veps(\varrho) = {1\over 2} \varrho^{\otimes 2}(G) + \varrho(V)\, .
\eqno\eqlbl\Efctnl
$$
The regular part, $F$, of $G$ does not contribute to the limit. 
We list a few examples.

\medskip
\noindent
{\bf IVa.} {\it Real symmetric, complex Hermitian, 
		and quaternionic self-dual Hermitian matrices}

As mentioned in the introduction we have 
$\Lambda =  \RR$, $G(\lambda,\eta) = - \ln |\lambda - \eta|$, 
$\beta = 1,2,4$, and $\kappa = 1,2,2$, respectively. 
Our electrostatic variational principle
(VP) for (\Efctnl) then becomes the VP of Boutet de Monvel et al. 
[\boutetetal], but here with a slightly wider class of potentials $V$. 
In particular, for $\beta =1$ we can allow continuous $V$ with 
$V(\lambda) \sim (1+\eps)\ln|\lambda|$ asymptotically, as compared to
H\"older continuous $V$ with $V(\lambda) \sim (2+\eps)\ln|\lambda|$ 
in [\boutetetal]. The VP has been studied extensively in [\safftotikBOOK].
A unique minimizer is known to exist under certain regularity conditions
on $V$. Of course, for $V(\lambda) = |\lambda|^2$, the quadratic potential 
of the Gaussian ensembles, the minimizer of (\Efctnl) is given by 
Wigner's semicircle law (\sclaw). 

\medskip
\noindent
{\bf IVb.} {\it General complex matrices}

We have $\Lambda = \RR^2$, $G(\lambda,\eta) = - \ln |\lambda - \eta|$.
In this case our variational principle for (\Efctnl) 
generalizes the VP of [\boutetetal] to two-dimensional domains.
Under mild conditions on $V$, and in particular for 
all our examples, it can be shown that the minimizer is 
unique.

We consider only the Gaussian ensemble with $\kappa V(M) = M^\dagger M$
in (\wishartmeas), whence $V(\lambda) = |\lambda|^2/2$ in (\hamiltA),
and $\beta =2$ in (\jEVmeas). The minimizer of (\Efctnl) is given by 
$$
d\varrho_0 = \pi^{-1}\chi_{\rm B_1}^{\phantom{n}}(\lambda)\,d\lambda\, ,
\eqno\eqlbl\diskmeas
$$  
where $\chi_{\rm B_1}^{\phantom{n}}(\lambda)$ is the indicator
function of the unit disk B$_1$ in $\RR^2$. This result can 
also be obtained from Ginibre's exact finite $N$ formula, see [\ginibre].  

\medskip
\noindent
{\bf IVc.} {\it Complex normal matrices}

We have $\Lambda = \RR^2$, $G(\lambda,\eta) = - \ln |\lambda - \eta|$,
and $\beta =2$. Consider first (\wishartmeas) with
$\kappa V(M)=\ln(1 + M^\dagger M)^{1+1/N}$. Then in (\hamiltA) we have
$V(\lambda) = - \ln [\pi\rho_C^{\phantom{n}} (|\lambda|)]^{1/2}$, 
where $\rho_C^{\phantom{n}}(\xi) = \pi^{-1}(1+\xi^2)^{-1}$ is the density 
of the Cauchy distribution, and $F$  is replaced by $V$.
With these identifications $(C5)$ is violated, but (\jEVmeas) is  
well defined for all $\beta >1$, and the minimizer of the 
electrostatic energy functional is found to be 
$$
d\varrho_0 = \pi^{-1}(1+|\lambda|^2)^{-2}d\lambda \, .
\eqno\eqlbl\stereogrmeas
$$  

The measure (\stereogrmeas) has geometrical significance. 
Recall that $|J|^2(\lambda) = 4/(1+|\lambda|^2)^{2}$ is the Jacobian 
of the stereographic projection map $\SS^2 \to \RR^2$, arranged such
that the equator of $\SS^2$ coincides with the unit circle in $\RR^2$.
Therefore, (\stereogrmeas) is the stereographic projection onto the 
Euclidean plane of the uniform probability measure on ${\SS}^2$. 
Also the finite-$N$ measure (\jEVmeas), with (\hamiltA) 
specified as above, is itself a stereographic projection
onto Euclidean space of a canonical ensemble 
measure of $N$ point charges in the two-sphere ${\SS}^2$. 
For $\beta =2$, this {\it spherical ensemble} is  given by (\jEVmeas) 
with Hamiltonian 
$$
H^{(N)}(\lambda_1,...,\lambda_N)
 = - \sum_{1\leq j<k\leq N} \ln |\lambda_j - \lambda_k| \, ,
\eqno\eqlbl\Shamilt
$$
where $\lambda_j \in \SS^2$, $|\lambda_j - \lambda_k|$ is the
chordal distance on $\SS^2$, and $d\lambda_k$ in (\jEVmeas) now 
means uniform measure on $\SS^2$. 
(If $\beta \neq 2$, a non-constant one-particle potential has to be added
to $H^{(N)}$). Our theorem applies directly to this ensemble 
on $(\SS^2)^{\times N}$. We arrive at 
(\stereogrmeas) by taking the limit $N\to \infty$ on the sphere,
arguing that the minimizer is a constant,
and projecting the result onto the Euclidean plane. 

In a sense, the spherical ensemble is the 
counterpart on ${\SS}^2$ to Dyson's circular ensembles on ${\SS}^1$,
although these ensemble are related to random matrices in a different
manner. The statistical mechanics of the spherical ensemble has been studied 
in some detail in [\caillol,\forrjanco], using exact algebraic techniques. 
The spherical ensemble is also related to J.J. Thomson's celebrated 
problem of determining the minimum energy configuration of $N$ point 
charges on $\SS^2$ [\jjthomson], which recently has received much 
attention in physics [\altschuleretal], topology and Knot theory 
[\kusnersullivan]. 

As a second example, consider entries restricted by
$\|M\|_\infty \leq 1$ for every $M$, with uniform distribution otherwise.
This corresponds to  
$V(M)= \lim_{t\to\infty}(1+\tanh[t(I- MM^\dagger)])^{-1}$. 
Alternatively, we can choose $\Lambda = {\rm B}_1$ and $V = 0$ in (\jEVmeas). 
The unique minimizer of the corresponding variational principle 
is the electrostatic  charge distribution on a 
circular perfect conductor. It is well known that any
surplus charge accumulates at the ``surface,'' 
 i.e., the minimizer is given by a Dirac mass 
concentrated uniformly on the boundary of the unit disk, 
$$
d\varrho_0 = \delta_{\SS^1}^{\phantom{n}}(dx)\, .
\eqno\eqlbl\circlemeas
$$  
With the help of our corollary this gives us the following. 

\smallskip\noindent
{\bf Proposition:} {\it Let $\bigl\langle f \bigr\rangle_N$ be
defined as in (\average), the summation running over
the eigenvalues, restricted to B$_1\subset\CC$,
of a complex normal $N\times N$ random matrix whose free
entries are uniformly distributed otherwise. Then, in probability,  
$\bigl\langle f \bigr\rangle_N  \to ({2\pi})^{-1}\int_0^{2\pi}
f\left(e^{i\varphi}\right)d\varphi$ as $N\to\infty$.}
\medskip

This result is worth rephrasing in less technical terms. 
As far as averages over the spectrum are concerned, 
an infinite normal random matrix with eigenvalues 
in the unit disk and independent entries uniformly 
distributed otherwise, is almost surely equivalent to some 
infinite unitary matrix.  

\medskip
\noindent
{\bf IVd.} {\it Real quaternion matrices}

We consider only the Gaussian ensemble. The joint probability density  
is given by (\specmeasQ) with $\Lambda = \RR^2$, $\beta =2$, $\kappa =
1$, [\ginibre,\mehtaBOOK]. The limiting eigenvalue distribution now 
minimizes the slightly more complicated electrostatic energy functional 
$$
\veps(\varrho) = {1\over 2} \varrho^{\otimes 2}
\(- \ln |\lambda - \eta| -\ln |\lambda - \eta^*| +|\lambda|^2 +|\eta|^2\)\, ,
\eqno\eqlbl\rqEfctnl
$$
where $\eta^*$ is the mirror image of $\eta$ with respect to the 
real axis. The  minimizer of (\rqEfctnl) is, once again, unique and
given by the measure (\diskmeas).

\medskip

{\bf ACKNOWLEDGEMENT.} 

We thank the referee for a careful reading of the manuscript,
and L. Pastur and P. Zinn-Justin for valuable comments.
The work of M.K. was supported by NSF Grant \# DMS 96-23220. 

\biblio

\bye